\documentclass[twocolumn,superscriptaddress,floatfix,preprintnumbers]{revtex4}
\usepackage{graphics,amssymb,amsmath,epsfig,color}
\usepackage{graphicx}
\usepackage[latin9]{inputenc}
\usepackage[active]{srcltx}
\usepackage{color}
\usepackage[english]{babel} 
\usepackage{amsmath}
\usepackage{amssymb}
\usepackage{graphicx}
\usepackage{csquotes}  

\usepackage{hyperref}
 
\hypersetup{
 colorlinks,linkcolor=red,citecolor=blue}
\usepackage{breakurl}
\usepackage{lipsum}

\makeatletter
\newcommand*{\balancecolsandclearpage}{%
  \close@column@grid
  \twocolumngrid
}

\usepackage{amsfonts}
\makeatother

\begin{document}

\title{Photonic reservoir control for Kerr soliton generation \\in strongly Raman-active media}

\author{Zheng Gong}
\affiliation{Department of Electrical Engineering, Yale University, New Haven, CT 06511, USA}
\author{Ming Li}
\affiliation{Department of Optics and Optics Engineering, University of Science and Technology of China, Hefei, Anhui 230026, China}
\author{Xianwen Liu}
\affiliation{Department of Electrical Engineering, Yale University, New Haven, CT 06511, USA}
\author{Yuntao Xu}
\affiliation{Department of Electrical Engineering, Yale University, New Haven, CT 06511, USA}
\author{Juanjuan Lu}
\affiliation{Department of Electrical Engineering, Yale University, New Haven, CT 06511, USA}
\author{Alexander Bruch}
\affiliation{Department of Electrical Engineering, Yale University, New Haven, CT 06511, USA}
\author{Joshua B. Surya}
\affiliation{Department of Electrical Engineering, Yale University, New Haven, CT 06511, USA}
\author{Changling Zou}
\affiliation{Department of Electrical Engineering, Yale University, New Haven, CT 06511, USA}
\affiliation{Department of Optics and Optics Engineering, University of Science and Technology of China, Hefei, Anhui 230026, China}
\author{Hong X. Tang}
\affiliation{Department of Electrical Engineering, Yale University, New Haven, CT 06511, USA}
\affiliation{Corresponding author: hong.tang@yale.edu}

\begin{abstract}
Microcavity soliton\textcolor{black}{s} enable miniaturized coherent frequency comb sources. \textcolor{black}{However, the formation of microcavity solitons} can be disrupted by stimulated Raman scattering (SRS), particularly in the emerging crystalline microcomb materials with high Raman gain. Here, we propose and implement \textcolor{black}{dissipation control}---tailoring the energy dissipation of selected cavity modes---to purposely raise/lower the threshold of Raman lasing in a strongly Raman-active lithium niobate microring resonator, and realize on-demand soliton mode-locking or Raman lasing. Numerical simulations are carried out to confirm our analyses and agree well with experiment results. \textcolor{black}{Our work demonstrates an effective approach to address strong SRS for microcavity soliton generation.} 
\end{abstract}

\maketitle
Microcavity solitons \cite{temoralsoliton} are miniaturized coherent frequency comb sources that have promising applications from time-frequency metrology to advanced communications \cite{solitonrev2}. However, their formation inside the cavity can be strongly perturbed by SRS \cite{Okawachi:17,Wang18} originating from inelastic scattering of photons by lattice phonon modes \cite{AGRAWAL2013295}. When the pump field energy is above a threshold level, Raman lasing is initiated and interferes with the four-wave mixing (FWM) process \cite{Griffith:16,doi:10.1021/acsphotonics.7b01254,Choi:18,Jung:19}, disrupting soliton formation \cite{Okawachi:17}. This is of particular concern in comb materials with strong Raman gain such as silicon \cite{Griffith:16}, diamond \cite{Latawiec:15}, GaP \cite{wilson2018integrated}, AlN \cite{Liu:17,Jung:19}, and lithium niobate (LN) \cite{Yu:19}. The latter three are emerging crystalline materials that hold great potential for on-chip comb self-referencing \cite{He:19,Jung:13,Bruch:19,Gong:18,wilson2018integrated} due to their simultaneous $\chi^{3}$ and $\chi^{2}$ optical nonlinearities. 
 
Methods to mitigate SRS include reducing the microresonator size \cite{Okawachi:17}, and orientating field polarization along the proper crystal axis \cite{solitonxcut}, which serve to reduce the spectral overlap between soliton-forming modes and dominating Raman gain. However, these solutions impose limits on device geometry and may affect the extent of dispersion control. On the other hand, optical microresonators can also be considered an \enquote{open} system driven by an external field, while dissipating energy either through the cavity's internal losses or by coupling to external waveguides. Thus SRS could also be manipulated from the perspective of \textcolor{black}{dissipation control}, through the adjustment of external coupling rates of the Raman mode with respect to soliton forming cavity modes. \textcolor{black}{Along this line of thinking, auxiliary microrings \cite{selectivelyQ/12.811613}, engineered pulley waveguide coupler \cite{PulleyQ:19}, and scattering centers \cite{selectivelyQ/12.811613,scattering2,scatteringcenter} have been proposed to modify the loss of cavity modes.}

In this Letter, we demonstrate \textcolor{black}{dissipation control} in the photonic domain using a microresonator formed on thin film LN, a highly Raman-active material,  to suppress Raman lasing and allow soliton comb generation. By exploiting a self-interference coupling structure, the external coupling rates among cavity modes are controlled to raise the Raman lasing intracavity pump mode threshold energy above the one needed for a single soliton, thus securing the soliton state. An analytical model is established to estimate and compare the intracavity pump mode threshold energies of the Raman lasing and Kerr soliton state. Also, numerical simulations are carried out to study the cavity dynamics. Both analytical and numerical results are consistent with experiment\textcolor{black}{al} observations.

\begin{figure}
\begin{centering}
\includegraphics{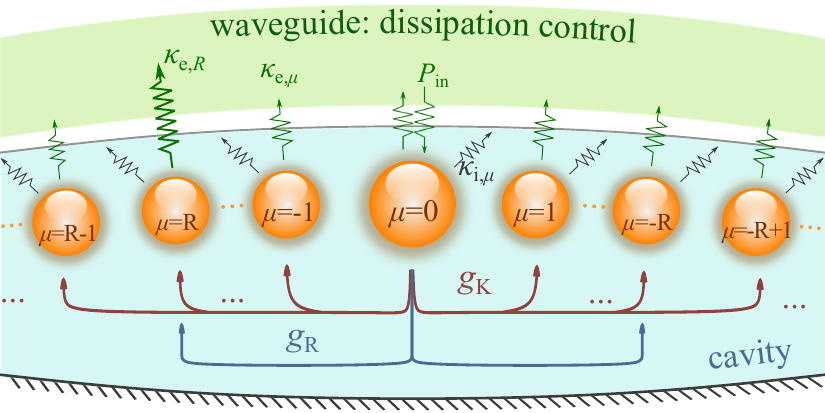}
\par\end{centering}
\caption{\label{fig:1} Bubbles: soliton-forming modes in the optical microresonator (blue-colored region), numbered with respect to the pump mode. Red arrowed lines: Kerr nonlinear coupling with a rate of $g_\mathrm{k}$. Blue arrowed lines: $1^\mathrm{st}$ SRS from the pump mode to the Stokes mode $\mathrm{R}$ with a rate of $g_\mathrm{R}$ and subsequent anti-SRS at mode $-\mathrm{R}$. \cite{AGRAWAL2013295}. Gray coils: cavity intrinsic dissipation rate ($\kappa_\mathrm{i,\mathrm{\mu}}$). Green coils: cavity-waveguide coupling rate ($\kappa_\mathrm{e,\mathrm{\mu}}$).}
\label{fig-1}
\end{figure}

A conceptual representation of \textcolor{black}{dissipation control} in a Raman-active optical microresonator is shown in Fig. \ref{fig:1}. The $\mu$-th soliton-forming mode dissipates energy through intrinsic loss channels at $\kappa_{\mathrm{i},\mu}$ and the external coupling waveguide at $\kappa_{\mathrm{e},\mu}$. In addition to FWM within the soliton-forming mode family, strong Raman effect will also transfer pump mode energy to the Stokes modes which could lead to Raman lasing, but can be suppressed by enhancing $\kappa_\mathrm{e,R}$. Specifically, we consider the $1^\mathrm{st}$ order lasing threshold of the dominant Raman mode in a microring \cite{Liu:17}, defined as the minimum required intracavity pump mode energy to initiate the Raman lasing  and given by (see Supplemental Material \cite{Supplement})
\begin{eqnarray}
\varepsilon_\mathrm{R,th}=\frac{\hbar\omega_\mathrm{p}\gamma_\mathrm{R}\kappa_\mathrm{R}}{4g^2_\mathrm{R}}[1+(\frac{2\delta_\mathrm{R}}{\gamma_\mathrm{R}})^2] \label{1}
\end{eqnarray}
where $\gamma_\mathrm{R}$ is the \textcolor{black}{full-width at half-maximum (FWHM)} of the Raman gain, $\kappa_\mathrm{R}=\kappa_\mathrm{i,R}+\kappa_\mathrm{e,R}$ is the total cavity decay rate of the Stokes mode ($\mathrm{\mu}=\mathrm{R}$, in Fig. \ref{fig:1}), $g_\mathrm{R}$ denotes the Raman coupling rate and $\delta_\mathrm{R}=\Omega_\mathrm{R}-(\omega_\mathrm{p}-\omega_\mathrm{R})$ represents the Raman gain detuning with $\omega_\mathrm{p}$ and $\omega_\mathrm{R}$ denoting pump and Stokes light angular frequencies. Note that the onset of the first order Raman lasing will result in a clamped pump mode energy at $\varepsilon_\mathrm{R,th}$. 

Under pure Kerr effect and ignoring dispersion of $3^{\mathrm{rd}}$ and above, the minimal intracavity pump mode energy required by a single soliton with an FWHM of $\gamma_\mathrm{S}$ can be approximated as \cite{temoralsoliton}
\begin{align}
\begin{split}
\varepsilon_\mathrm{S,\mathrm{th}}\simeq \frac{\hbar\omega_{\mathrm{p}}D_2}{4g_{\mathrm{K}}}[1+2(\frac{\kappa D_{1}}{\gamma_\mathrm{S}D_{2}})^2], \label{2}
\end{split}
\end{align}
\textcolor{black}{where $D_1$ represents the free spectral range ($D_1=2\pi{\mathrm{FSR}}$),} $D_2>0$ denotes the $2^{\mathrm{nd}}$ dispersion, and $g_{\mathrm{K}}=(\hbar\omega^2_0cn_2)/(n^2_0V_\mathrm{eff})$ is the Kerr nonlinear coupling rate with $c$, $n_2$, $n_0$, and $V_\mathrm{eff}$ being the speed of light in vacuum, nonlinear refractive index, effective refractive index, and mode volume respectively \cite{temoralsoliton}. In this analytical model, a single $\kappa=\kappa_{\mathrm{e}}+\kappa_{\mathrm{i}}$ is used to describe the total cavity decay rate for all the optical modes, while mode-dependent coupling rates are included in the numerical model \cite{Supplement}. The leading term in Eq. (\ref{2}) represents the energy of the frequency component at $\omega_\mathrm{p}$ of the pulse, and the second term signifies the energy of the \textcolor{black}{continuous wave (CW)} background.

Raman lasing can be suppressed by raising the Raman gain detuning $\delta_\mathrm{R}$ \cite{Okawachi:17}, causing an increase in $(\frac{\delta_\mathrm{R}}{\gamma_\mathrm{R}})^2$ and threshold. However, adjusting $\delta_\mathrm{R}$ alone imposes a lower limit on the soliton comb repetition rate \cite{Okawachi:17,solitonrev2}. On the other hand, if one manages to increase $\kappa_\mathrm{e,R}$ by engineering the cavity's external coupling configuration, then the restraint on the ring geometry will be largely relaxed.

\begin{figure}
\begin{centering}
\includegraphics{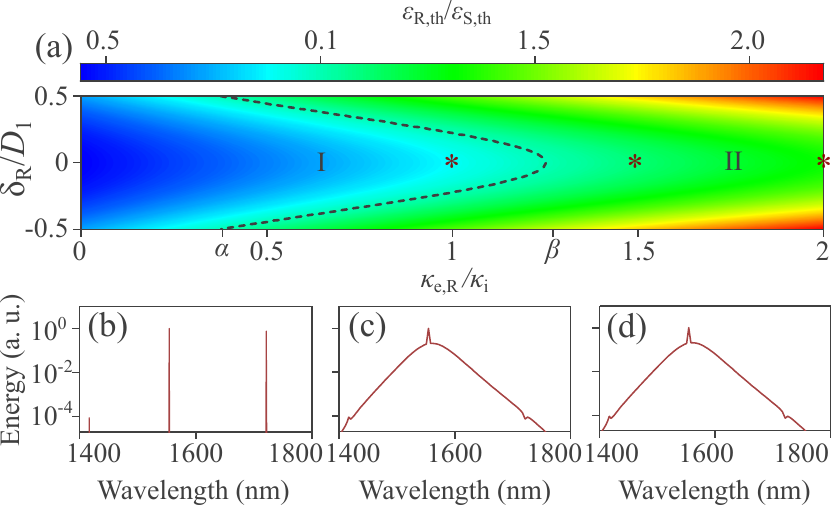}
\par\end{centering}
\caption{\label{fig:2} (a) Threshold ratio  $\varepsilon_\mathrm{R,th}/\varepsilon_\mathrm{S,th}$ as the Raman mode detuning $\delta_\mathrm{R}/D_{1}$ and external coupling ratio $\kappa_\mathrm{e,R}/\kappa_\mathrm{i}$ are varied. The dashed curve seperates the $\varepsilon_\mathrm{R,th}<\varepsilon_\mathrm{S,th}$ (I) and $\varepsilon_\mathrm{R,th}>\varepsilon_\mathrm{S,th}$ (II) regimes. (b-d) Simulated intracavity spectra for Raman lasing and soliton combs at $\delta_\mathrm{R}/D_{1}=0$ and $\kappa_\mathrm{e,R}/\kappa_\mathrm{i}=$ 1, 1.5 and 2 marked by asterisks in (a) respectively. Insets show the temporal profiles. For (a-d), $\kappa_\mathrm{i}/(2\pi)=\kappa_\mathrm{e}/(2\pi)=235$\,MHz, $D_2/\kappa_\mathrm{i}=8.6\times10^{-2}$. \textcolor{black}{The soliton self-frequency shifts in (c, d) are 0.9\,THz.}}
\end{figure}

To illustrate raising $\varepsilon_\mathrm{R,th}$ above $\varepsilon_\mathrm{S,th}$ with increased $\kappa_\mathrm{e,R}$, the calculated $\varepsilon_\mathrm{R,th}/\varepsilon_\mathrm{S,th}$ under different Raman detunings and external coupling rates are presented in Fig.~\ref{fig:2}. Here, we attempt to secure a single soliton, with a target FWHM of 6.5\,THz as an example, in a cavity where Raman gain FWHM is larger than the microring FSR, with $\gamma_\mathrm{R}/D_1=1.3$, chosen to reflect our actual device parameters. Both the Raman and Kerr coefficients are set using corresponding literature values of LN. To gain insight into the complex dynamics of the coupled nonlinear processes, only $\kappa_\mathrm{e,R}$ of the Stokes mode is varied while $\kappa_\mathrm{i}$ and $\kappa_\mathrm{e}$ remain unchanged for the other modes, and the calculation of $\varepsilon_\mathrm{S,th}$ based on Eq.~(\ref{2}) ignores the variation of $\kappa_\mathrm{e,R}$ at the Stokes mode which is far away from the pump mode. The numerical model presented later incorporates mode-dependent coupling rates, and the main conclusions drawn from our simplified model remain valid.

Fig.~\ref{fig:2}(a) delineates \textcolor{black}{separate regimes for Raman threshold being lower/higher than the soliton state's (I/II).} For small external coupling rate at the Stokes mode ($\kappa_\mathrm{e,R}/\kappa_\mathrm{i}$\,$<$\,$\alpha$), the Raman lasing threshold is always lower than that of soliton state, even if the Raman gain peak is optimally detuned from adjacent soliton-forming modes (in the middle between two modes). Here, $\alpha$\,$=$\,$4g^2_\mathrm{R}\varepsilon_\mathrm{S,th}/[\hbar\omega_\mathrm{p}\kappa_{\mathrm{i}}(\gamma_\mathrm{R}+{D_{1}^2}/{\gamma_\mathrm{R}})]-1$ is a material- and device-dependent dimensionless parameter. At increased Stokes mode external coupling rate ($\alpha<\kappa_\mathrm{e,R}/\kappa_{\mathrm{i}}<\beta=[1+({D_{1}}/{\gamma_\mathrm{R}})^2]\alpha$), the Raman lasing threshold can be elevated above the soliton state's when there is sufficient detuning between the Raman gain peak and adjacent modes.

By further increasing $\kappa_\mathrm{e,R}$ ($\kappa_\mathrm{e,R}/\kappa_\mathrm{i}$\,$>$\,$\beta$), the soliton state is favored over Raman lasing for all possible Raman-Stokes detunings. Numerical simulations carried out at several representative locations marked in the parameter space of Fig.~\ref{fig:2}(a) suggest that the single soliton state can be maintained, even when the Raman gain peak overlaps the Stokes mode. As examples, the simulated intracavity spectra for $\varepsilon_\mathrm{R,th}/\varepsilon_\mathrm{S,th}$\,$=$\,0.9, 1.1 and 1.3 are presented in Fig.~\ref{fig:2}(b-c).

\begin{figure}
\begin{centering}
\includegraphics{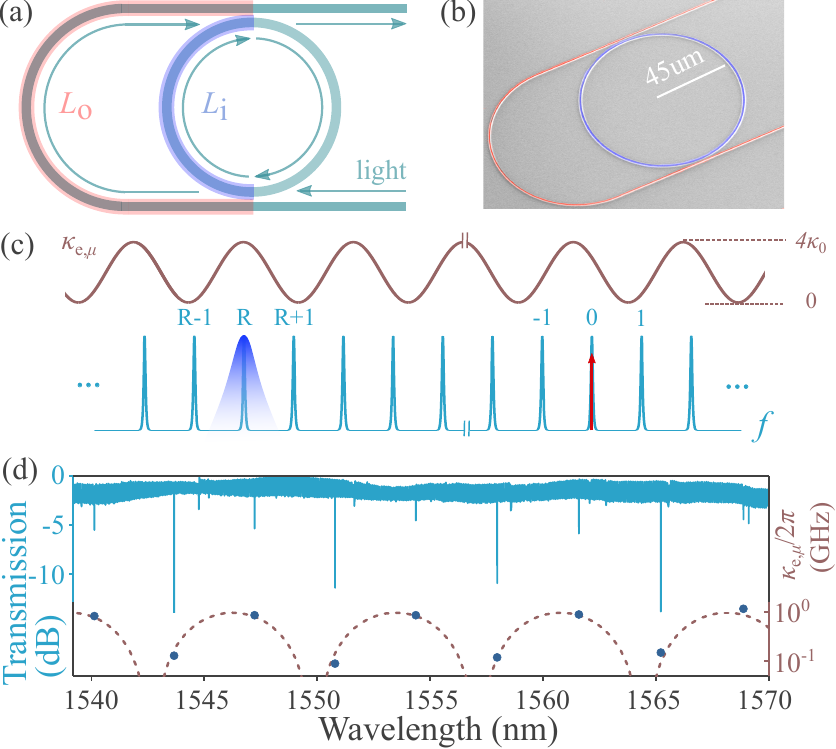}
\par\end{centering}
\caption{\label{fig:3} (a) Schematic of a self-interferenced microring. Purple (pink) shaded line: the inner (outer) arm of the interferometer, with its length denoted as $L_{\mathrm{i(o)}}$. (b) False-color SEM of the U-ring.  (c) Schematic illustration of modulated $\kappa_{\mathrm{e},\mu}$ (brown line) to suppress Raman lasing. Cyan peaks: soliton-forming modes. Red arrow marks the pump mode. Blue shaded profile: the Raman gain. (d) Cyan line: measured TE-transmission of (b). Dashed brown line: predicted $\kappa_{\mathrm{e}}$ curve from Eq. (\ref{3}). \textcolor{black}{Cyan dots: extracted $\kappa_{\mathrm{e},\mu}$ from the measured transmission by fitting the linewidths of the resonances.}} 
\end{figure}

\begin{figure*}
\begin{centering}
\includegraphics{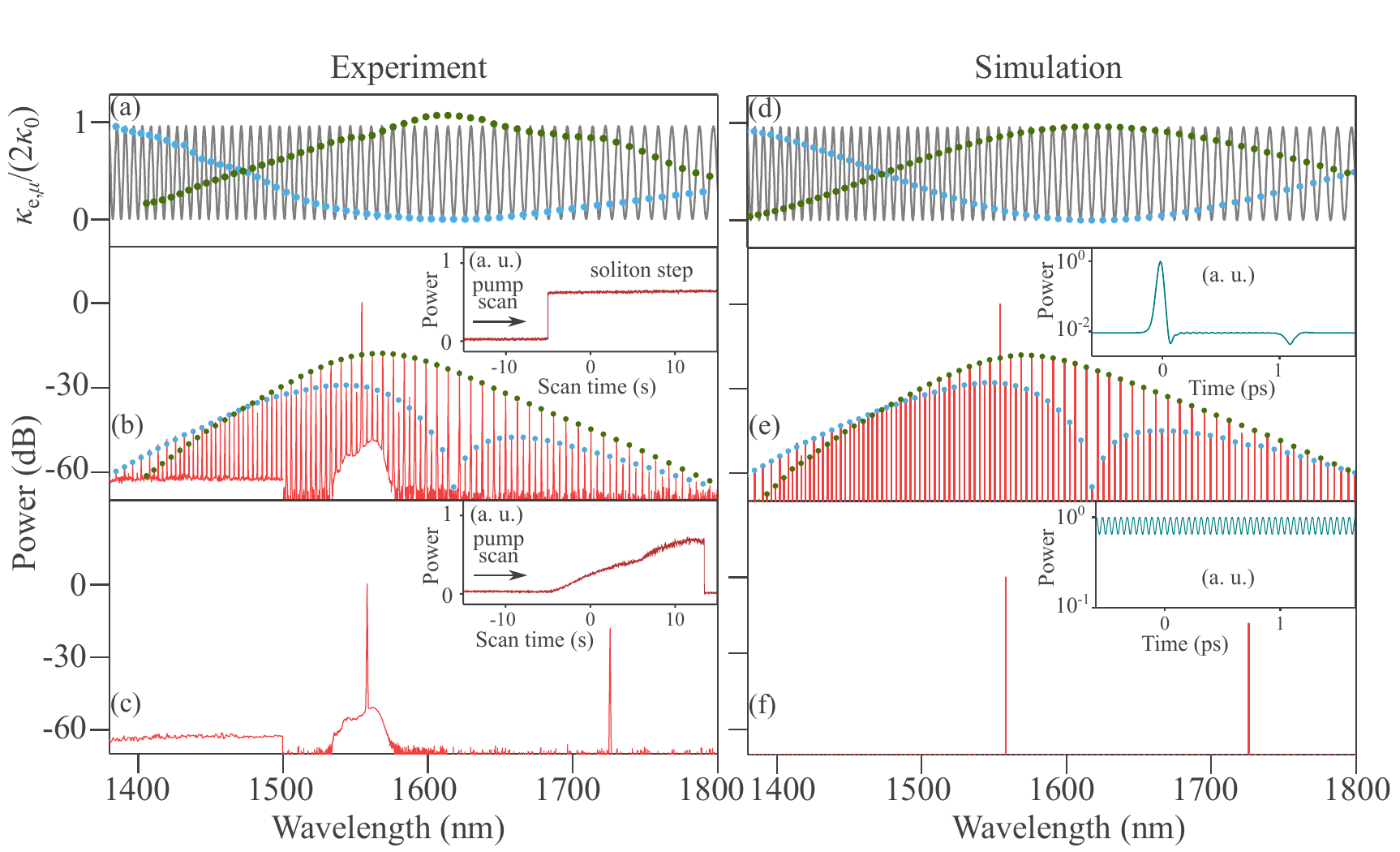}
\par\end{centering}
\caption{\label{fig:4} (a) External coupling rate $\kappa_{\mathrm{e},\mu}$, marked alternately with cyan/green dots, are extracted from the corresponding cyan/green dotted curve tracing the measured comb line powers in panel (b). The gray line in the background is the predicted continuous $\kappa_{\mathrm{e}}$ curve. The deviation of the extracted external couplings from the theoretically calculated may be ascribed to local mode-crossings which affects the extraction of $\kappa_{\mathrm{e},\mu}$. (b) The measured and normalized soliton comb output spectrum with an FSR of $\sim$\,445.7\,GHz, \textcolor{black}{under on-chip pump power of 200 mW}. Cyan/green dotted curve: the spectral envelope. Inset: comb power under red-to-blue pump scan (0.25\,GHz/s). (c) The measured and normalized $1^\mathrm{st}$ Raman lasing spectrum. Inset: Stokes light power trace under the same scan for the inset of (b). (d) Cyan/green dots: \textcolor{black}{$\kappa_{\mathrm{e},\mu}$ for the simulation, calculated by Eq.~\ref{3} with $\kappa_{0}/(2\pi)$\,$=$\,240\,MHz and $c/(L_\mathrm{o}n_\mathrm{o}-L_\mathrm{i}n_\mathrm{i})$\,$=$\,900\,GHz.} Gray line is the same to (a). (e, f) The simulated and normalized soliton comb and Raman lasing spectra on the output waveguide. Cyan/green dotted curve: spectral envelope imposed by $\kappa_{\mathrm{e},\mu}$ (cyan/green dots) in (d). Insets: intracavity temporal profiles. \textcolor{black}{The time interval between the soliton and the fed-back pulse is determined by the path difference, $(L_\mathrm{o}n_\mathrm{o}-L_\mathrm{i}n_\mathrm{i})/c$\,$=$\,1.1\,ps.}}
\end{figure*}

To implement the \textcolor{black}{dissipation control} concept described above, it is critical that the external coupling rate of the Raman modes can be engineered. For the purpose of tuning the external coupling rates of selected modes, the photonics community have established approaches including the use of a pulley waveguide \cite{Hosseini:10,Moille:19,PulleyQ:19} and auxiliary microring \cite{Miller:15,2ringmodecross,selectivelyQ/12.811613}. Here we adopt a self-interference structure (Fig. \ref{fig:3}(a)) which offers high dynamic tuning of $\kappa_\mathrm{e}$, integrability with local heater/electrode for fine tuning \cite{Chen:07,Wan:18}, as well as extendability for the suppression of multiple Raman lines. \textcolor{black}{The scanning electron micrograph (SEM)} image of our device is shown in Fig. \ref{fig:3}(b), where the external waveguide point-couples twice with the microring (referred to U-ring hereafter). The net external coupling rate of $\mu$-th mode $\kappa_\mathrm{e,\mu}$ is given by \cite{Supplement}
\begin{eqnarray}
\kappa_\mathrm{e,\mu}=2\kappa_\mathrm{0}(1+cos\phi_{\mu}) \label{3}
\end{eqnarray}
where $\kappa_\mathrm{0}$ is the coupling rate at each of the two identical microring-waveguide couplers, and $\phi_{\mu}=\frac{\omega(\mu)}{c}(L_\mathrm{o}n_\mathrm{o}-L_\mathrm{i}n_\mathrm{i})$ is the mode-dependent phase difference between the two arms with $L_\mathrm{i(o)}$ and $n_\mathrm{i(o)}$ denoting the length and effective index of the inner (outer) arm of the $\mu$-th mode. The schematic in Fig. \ref{fig:3}(c) illustrates the modulation of external coupling rates around the Stokes mode $\mu$\,$=$\,$\mathrm{R}$. The Stokes mode can be aligned to the peak of $\kappa_\mathrm{e}$ curve by using a modulation period larger than the Raman gain FWHM and selecting proper pump wavelength. \textcolor{black}{In the experiment, we utilized a fast modulation ($L_\mathrm{o}$\,$\approx$\,$2L_\mathrm{i}$) to provide higher probability of finding a suitable pump mode within the EDFA bandwidth.}

Our microring resonator is patterned from a z-cut LN thin film \cite{Gong:19,Lu:19} that exhibits anomalous dispersion $D_2/(2\pi)$\,$\approx$\,$20$\,MHz and possesses strong Raman gain $g_{\mathrm{R}}/g_{\mathrm{k}}$\,$\approx$\,$1.3$\,$\times$\,$10^{6}$ with a linewidth of $\gamma_{\mathrm{R}}$\,$\approx$\,$1.3D_{1}$. Without \textcolor{black}{dissipation control}, the SRS is dominant and leads to $1^\mathrm{st}$ order Raman lasing at the mode $\mu$\,$=$\,-42, corresponding to the dominant E(LO$_8$) Raman mode \cite{Liu:17,Supplement}. The transmission of the U-ring is presented in Fig.~\ref{fig:3} (d), where the extracted external coupling rates are plotted against the values predicted by Eq.~(\ref{3}) with $\kappa_\mathrm{0}/(2\pi)$\,$=$\,$1.1\kappa_\mathrm{i}/(2\pi)$\,$=$\,240\,MHz and a modulation period of 2.02FSR. To estimate $\varepsilon_\mathrm{S,th}$, we consider a $sech^2$-weighted external coupling rate $\kappa_{\mathrm{e}}=[\sum{sech^2(\mu/N)\kappa_{\mathrm{e},\mu}}]/[\sum{sech^2(\mu/N)}]$ which factors in $\kappa_\mathrm{e,\mu}$ being modulated across the soliton-forming modes and phenomenologically represents a collective soliton external coupling rate, where $N\cdot\mathrm{FSR}$ indicates the bandwidth of interest. \textcolor{black}{Under this simplification, $\varepsilon_\mathrm{S,th}$ can be estimated by Eq.~(\ref{2}), 1.2 times higher than the numerically simulated value considering mode-dependent external coupling rates \cite{Supplement}.}

To suppress Raman lasing in favor of soliton generation, we chose to pump a TE$_{00}$ mode at 1554.4\,nm whose Stokes mode sees a high external coupling rate $\kappa_\mathrm{e,R}=3.7\kappa_\mathrm{0}$ as inferred from Fig. \ref{fig:3}(d). The estimated Raman lasing threshold is raised above the targeted soliton state threshold by $\varepsilon_\mathrm{R,th}/\varepsilon_\mathrm{S,th}$\,$=1.1$. As expected, the single soliton comb can be successfully generated in the experiment by scanning the pump along the red-side of the resonance \cite{He:19,Gong:19} until the soliton step shows up (Fig.~\ref{fig:4}(b) inset). The recorded output spectrum is shown in Fig. \ref{fig:4}(b), in which each comb line power is modulated by its external coupling rate, $P_\mathrm{S,\mu}\propto\varepsilon_\mathrm{S,\mu}\kappa_\mathrm{e,\mu}$, where $\varepsilon_\mathrm{S,\mu}$ is the ${\mu}$-th mode's intracavity energy. Assuming that the intracavity mode energy $\varepsilon_\mathrm{S,\mu}$ maintains a $sech^2$-profile, \textcolor{black}{$\kappa_\mathrm{e,\mu}$ can be separately extracted from the measured spectrum (Fig. \ref{fig:4}(a)).} The numerical simulation shown in Fig. \ref{fig:4}(e) verifies that the single soliton can be obtained under this configuration. \textcolor{black}{It is notable that, in a time domain picture, the intracavity soliton couples to the U-arm and is subsequently fed-back to the microring every roundtrip, introducing a weak fed-back pulse of less than 1\% intensity of the main pulse (inset of Fig.~\ref{fig:4}(e)). This time-domain picture is captured in the frequency domain by the modulated $\kappa_\mathrm{e,\mu}$ (Fig.~\ref{fig:4}(d)). Here a dip is observed due to the out-of-phase delay introduced by the chosen path difference $L_\mathrm{o}n_\mathrm{o}-L_\mathrm{i}n_\mathrm{i}$.} 

The U-ring switches to Raman lasing, with no solitons observed, when pumped at 1558\,nm (one FSR away from the previous setting). At this detuning, the Stokes mode sees a smaller $\kappa_\mathrm{e,R}= 0.4\kappa_\mathrm{0}$ (Fig. \ref{fig:3}(d)). As a result, the lasing threshold $\varepsilon_\mathrm{R,th}$ drops by a factor of 3.6 and leads to an estimated $\varepsilon_\mathrm{R,th}/\varepsilon_\mathrm{S,th}$\,=\,0.3. The recorded Raman lasing output spectrum is displayed in Fig. \ref{fig:4}(c). Only Stokes Raman optical power is recorded in the power time trace with no soliton steps observed (Fig. \ref{fig:4}(c) inset).  Numerically simulated output spectrum shown in Fig. \ref{fig:4}(f) confirms Raman lasing without the formation of the Kerr soliton. 

In conclusion, the use of \textcolor{black}{dissipation control} in a Raman-active microcavity to suppress Raman lasing for soliton generation was demonstrated. Theoretical analyses and numerical simulations suggest that the competition between intracavity Raman lasing and soliton formation can be controlled by systematically engineering the external coupling rates, where the final steady state relies on which state has lower threshold pump mode energy. The concept is implemented via a self-interference coupling structure. By pumping different modes along the modulation curve, soliton mode-locking and Raman lasing can be steered on-demand in a single device. This design concept can be extended to more complex coupling structures to suppress multiple Raman gain lines or stronger Raman effect \cite{Supplement}. Our work provides guidance to overcome challenges related to the competition between intracavity Raman lasing and soliton formation from the perspective of \textcolor{black}{dissipation control}, and could inspire future studies on regulating intracavity dynamics while multiple nonlinear processes are present.

\textbf{Acknowledgment} We acknowledge funding support from Defense Advance Research Project Agency.  H.X.T. acknowledges support from a Packard Fellowship in Science and Engineering. The authors thank Yong Sun, Sean Rinehart, Kelly Woods, and Michael Rooks for assistance in device fabrication.

\pagebreak
\widetext
\begin{center}
\textbf{\large Photonic reservoir control for Kerr soliton generation \\in strongly Raman-active media: supplementary material, Gong et al.}
\end{center}
\balancecolsandclearpage


This document provides supplementary information to the main manuscript titled \enquote{Photonic dissipation control for Kerr soliton generation in strongly Raman-active media}, including: the derivation of the numerical simulation model that incorporates both Raman and Kerr effects, derivation of the intracavity pump energy threshold for the 1$^{\mathrm{st}}$ order Raman lasing, the measurement of $g_{\mathrm{R}}$, the derivation of the net external coupling rates of the U-ring, experiment setup and process to access soliton state. Prospects are provided in the last section on further enhancing the system performance via cascaded self-interference scheme.

\setcounter{equation}{0}
\setcounter{figure}{0}
\setcounter{table}{0}
\setcounter{page}{1}
\makeatletter
\renewcommand{\theequation}{S\arabic{equation}}
\renewcommand{\thefigure}{S\arabic{figure}}
\renewcommand{\bibnumfmt}[1]{[S#1]}
\renewcommand{\citenumfont}[1]{S#1}

\section{Numerical model: Coupled mode equations of Raman and Kerr processes in a Microring }
Our simulation treats the Raman scattering as a second-order nonlinear process between two photonic modes and one Raman phonon mode, with the interaction Hamiltonian given by 
\begin{eqnarray}
\mathcal{H_{\mathrm{R}}} & = & \sum_{j,k,l}g_{\mathrm{R}}^{ijk}\left(a_{j}^{\dagger}a_{k}R_{l}+a_{j}a_{k}^{\dagger}R_{l}^{\dagger}\right),
\label{S01} 
\end{eqnarray}
where $a_{j,k}$ are the annihilation operator of the photonic modes,
$R_{l}$ is the annihilation operator of the Raman phonon mode, $g_{\mathrm{R}}$
is the Raman coupling rate. The subscripts $j,k,l$ denote the
angular momentum of these operators. Since the nonlinear interaction
requires that the photonic modes are phase-matched with the phonon
modes, we have 
\begin{align}
g_{\mathrm{R}}^{ijk} & \propto & \epsilon_{0}\chi_{\mathrm{R}}\int\int u_{x,j}\left(r,z\right)u_{x,k}\left(r,z\right)\boldsymbol{l_{0}}\delta\left(j-k-l\right),
\label{S02}
\end{align}
where $\chi_{\mathrm{R}}$ is the Raman scattering tensor, $u\left(r,z\right)$
is the mode distribution of the photonic mode. For the frequency
range considered here, we treat $g_{\mathrm{R}}^{ijk}$ as a constant value $g_{\mathrm{R}}$. 

Together with the interaction Hamiltonian of four-wave mixing, the
total Hamiltonian is written as 
\begin{eqnarray}
\mathcal{H} & = & \mathcal{H}_{\mathrm{sys}}+\mathcal{H}_{\mathrm{K}}+\mathcal{H}_{\mathrm{R}},
\label{S03}
\end{eqnarray}
where
\begin{eqnarray}
\mathcal{H}_{\mathrm{sys}} & = & \sum_{i}\omega_{a,i}a_{i}^{\dagger}a_{i}+\sum_{j}\omega_{R,j}R_{j}^{\dagger}R_{j},\\
\mathcal{H}_{\mathrm{K}} & = & g_{\mathrm{K}}\delta\left(i+j-k-l\right)\sum_{ijkl}a_{i}^{\dagger}a_{j}^{\dagger}a_{k}a_{l},
\label{S04}
\end{eqnarray}
where $g_{\mathrm{K}}$ is the Kerr nonlinear coupling strength \cite{temoralsoliton} and the phase-matching
condition is described by the delta function. Now, introducing the Fourier
transform of the operator $a_{i}$
\begin{align}
\mathit{E_{v}=\mathcal{F}_{v}}[a_{\mu}]=\frac{1}{\sqrt{N}}\sum_{\mu}a_{\mu}e^{2i\pi\mu v/N},
\end{align}\label{S05}
we have
\begin{align}
\mathcal{F}_{v}\Big[\sum_{k,l}a_{k}^{\dagger}a_{l}\delta\left(l-k-\mu\right)\Big]=\sqrt{N}\mathcal{F}[\overrightarrow{a}]^{\dagger}\mathcal{F}[\overrightarrow{a}],
\label{S06}
\end{align}
and
\begin{align}
\mathcal{F}_{v}\Big[\sum_{k,l,n}a_{k}^{\dagger}a_{l}a_{n}\delta\left(l+n-k-\mu\right)\Big]= N\mathcal{F}[\overrightarrow{a}]^{\dagger}\mathcal{F}[\overrightarrow{a}]\mathcal{F}[\overrightarrow{a}].\label{S07}
\end{align}
Now the dynamics of the mean fields of both the photonic modes and the phononic modes can be derived as follows
\begin{align}
\begin{split}
\frac{d}{dt}\overrightarrow{a}=M_{a}\overrightarrow{a}-g_{\mathrm{K}}N\mathcal{F}^{-1}\{ \mathcal{F}[\overrightarrow{a}]^{\dagger}\mathcal{F}[\overrightarrow{a}]\mathcal{F}[\overrightarrow{a}]\}\\-ig_{\mathrm{R}}\sqrt{N}\mathcal{F}^{-1}\{F[\overrightarrow{a}]\mathcal{F}[\overrightarrow{\mathrm{R}}]\}-ig_{\mathrm{R}}\sqrt{N}\mathcal{F}^{-1}\{F[\overrightarrow{a}]\mathcal{F}[\overrightarrow{\mathrm{R}}]^{\dagger}\}\\+\mathcal{E}_{\mathrm{P}},\label{S08}
\end{split}
\end{align}
\begin{align}
\frac{d}{dt}\overrightarrow{\mathrm{R}}=M_{\mathrm{R}}\overrightarrow{\mathrm{R}}-ig_{\mathrm{R}}\sqrt{N}\mathcal{F}^{-1}\left\{ F[\overrightarrow{a}]^{\dagger}\mathcal{F}[\overrightarrow{a}]\right\},\label{S09}
\end{align}
where $M_{a}=\{\chi_{-N}, \chi_{-N+1},...,\chi_{N-1},\chi_{N}\}$, $\chi_{j}=-i\delta_{j}-\kappa_{j}/2$ with $\delta_{j}$ and $\kappa_{j}$ denoting the detuning between cold cavity angular frequencies and equidistant $D_{1}-$spaced angular frequency grid and total decay rate respectively \cite{temoralsoliton}, and similar for $M_{\mathrm{R}}$. $\mathcal{E}_{\mathrm{P}}=\{0,0,...,\sqrt{\kappa_{e,0}\mathrm{P}_{\mathrm{in}}/(\hbar\omega_{\mathrm{P}})},...,0,0\}$ represents the driving strength under input power of $\mathrm{P}_{\mathrm{in}}$, with only the $(N+1)^{\mathrm{th}}$ element being non-zero. 

Usually, the decay rate of Raman phonons $\gamma_{\mathrm{R}}$ is much larger
than the cavity photonic decay rate $\kappa_{j}$. Thus, it is reasonable to adiabatically
eliminate the photon mode by setting the time derivative of $\overrightarrow{\mathrm{R}}$
to be zero, then we get
\begin{eqnarray}
\overrightarrow{\mathrm{R}}=ig_{\mathrm{R}}\sqrt{N}\mathcal{F}^{-1}\left\{ F[\overrightarrow{a}]^{\dagger}\mathcal{F}[\overrightarrow{a}]\right\} /M_{\mathrm{R}},\label{S10}
\end{eqnarray}
and by doing so, the dynamics of the nonlinear system can be solved with fast numerical codes.
 
\begin{figure}
\begin{centering}
\includegraphics[width=1\columnwidth]{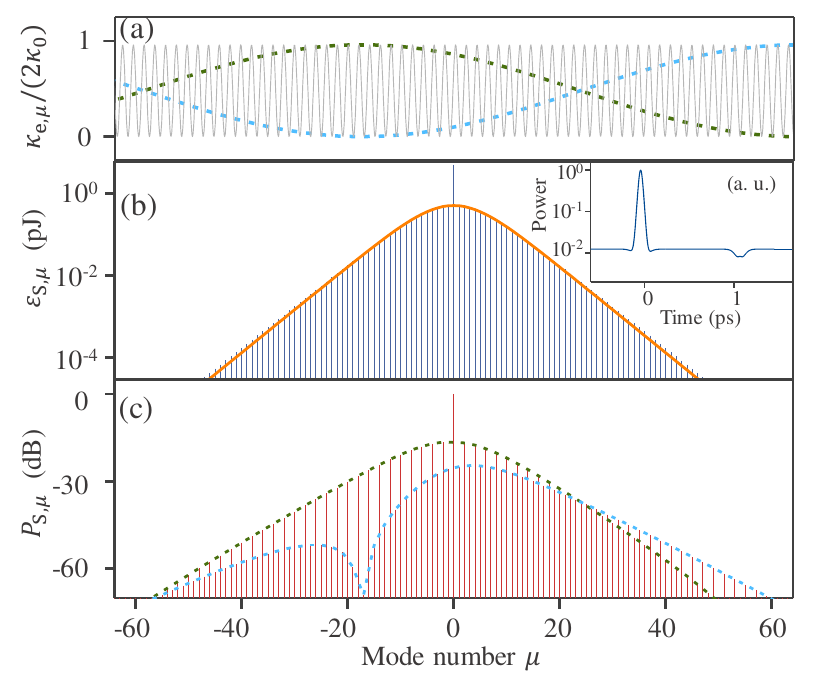}
\par\end{centering}
\caption{\label{fig:S1} (a) External coupling rates $\kappa_{\mathrm{e,\mu}}$ of the soliton forming modes used for the numerical simulation (Dashed green/blue line). Gray lines in the background are calculated from the theoretical continuous $\kappa_{\mathrm{e}}$ curve, same to Fig. 4(a) of the main text. (b) Simulated intracavity mode energy (in log-scale) of the maximally red-detuned soliton state with Raman gain turned off. Orange curve: $sech^{2}$-fitting curve, proportional to $sech^2(\mu/N)$ with N\,$=$\,8.3 ($\gamma_\mathrm{S}/(2\pi)$\,$=$\,6.5\,THz). Inset shows the temporal profile. (c) Simulated on-chip output optical spectrum of the U-ring. The blue/green dashed line shows the spectral envelope imposed by the group of $\kappa_{\mathrm{e,\mu}}$ (blue/green) in (a).}
\end{figure}

As a first step, we turn off the Raman effect and with  Kerr effect only to simulate the intracavity field under the modulated $\kappa_{\mathrm{e,\mu}}$ to evaluate soliton threshold energy $\varepsilon_\mathrm{S,th}$ and compare it to the result based on the simple model introduced in the main text using a mode-independent $sech^2$-weighted external coupling rate $\kappa_{\mathrm{e}}=[\sum{sech^2(\mu/N)\kappa_{\mathrm{e},\mu}}]/[\sum{sech^2(\mu/N)}]$.  Specifically, plugging in the same microring parameters used for the estimation of $\varepsilon_\mathrm{S,th}$ when pumping our device at $\sim$\,1554.4\,nm in the main text, the simulated soliton intracavity mode energy $\varepsilon_\mathrm{S,\mu}$ and corresponding microring output optical spectrum $P_\mathrm{S,\mu}$ are plotted in the Fig. \ref{fig:S1}. As can be seen, Fig. \ref{fig:S1}(b) validates the assumption made in the main text that the profile of $\varepsilon_\mathrm{S,\mu}$ still maintains the $sech^2$-shape. And the simulated $\varepsilon_\mathrm{S,th}$ (the total intracavity energy at the pump frequency in the soliton state near the end of its red-detuning range \cite{Yang:16,temoralsoliton}, shown in Fig.~\ref{fig:S1}(b)) turns out to be $\sim$\,4.9 pJ, which is $\sim$\,1.2 times less than the estimated value based on the constant external coupling rate model ( $\kappa_{\mathrm{e}}\simeq{2\kappa_\mathrm{0}}$) to achieve the same soliton bandwidth. The above comparison indicates the calculation of $\varepsilon_\mathrm{S,th}$ using the collective $sech^2$-weighted external coupling rate is a conservative estimation, when considering how much $\varepsilon_\mathrm{R,th}$ should be raised over $\varepsilon_\mathrm{S,th}$ for our device. \textcolor{black}{Compared with the case without dissipation control, i. e. $\kappa_{\mathrm{e}}/(2\pi)=\kappa_{0}/(2\pi)=$240\,MHz, the estimated threshold $\varepsilon_\mathrm{S,th}$ is approximately doubled.}

\begin{figure}
\begin{centering}
\includegraphics[width=1\columnwidth]{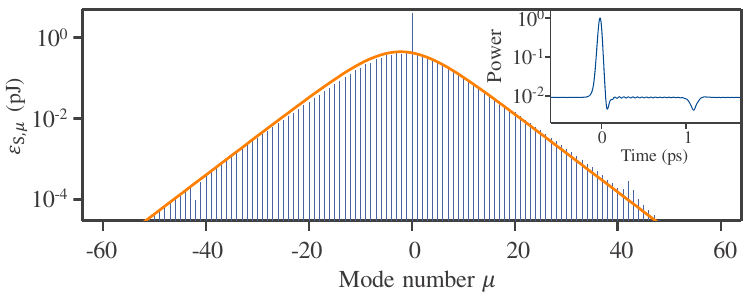}
\par\end{centering}
\caption{\label{fig:S2} The simulated intracavity mode energy (in log-scale) of the soliton state under $g_\mathrm{{R}}=1.1\times10^{7}$ along with $sech^{2}-$fitting (Orange curve). All the other parameters are identical to those used for the simulation in Fig. \ref{fig:S1}. Inset shows the temporal profile.}
\end{figure}
Next, we simulate the intracavity soliton spectrum when Raman effect is turned on, with $g_\mathrm{{R}}/(2\pi)=1.75$\,MHz set to the measured value of our device. The profile of $\varepsilon_\mathrm{S,\mu}$ is found to maintain the $sech^2$-shape as well (Fig. \ref{fig:S2}), which verifies the assumption made in the main text when extracting $\kappa_\mathrm{e,\mu}$ from the measured comb optical spectrum. And the intracavity profile used to extract $\kappa_\mathrm{e,\mu}$ in the main text (Fig. 4(a)) has a similar bandwidth to that shown in Fig. \ref{fig:S2}. Note that the simulated $\varepsilon_\mathrm{S,th}$ (the intracavity energy of the pump mode of the maximally red-detuned soliton state \cite{Yang:16,temoralsoliton}) now reduces to 4.2\,pJ from 4.9\,pJ in the above-mentioned pure Kerr case, which may be casused by the Raman-induced soliton self-frequency shift where energy transfers from high-frequency to low-frequency components \cite{PhysRevLett.116.103902,Yang:16}.

\section{Threshold for 1$^{\mathrm{st}}$ Raman lasing}

In absence of four wave mixing, the Hamiltonian for the intracavity 1$^{\mathrm{st}}$ Raman scattering from one dominant phonon mode is given by

\begin{align}
\begin{split}
\mathcal{H} = \omega_{a}a^{\dagger}a+\omega_{b}b^{\dagger}b+\omega_{\mathrm{R}}R^{\dagger}R+g_{\mathrm{R}}\left(b^{\dagger}R^{\dagger}a+bRa^{\dagger}\right)\\+i\sqrt{\kappa_{\mathrm{e},a}}\varepsilon_{\mathrm{p}}\left(a^{\dagger}e^{-i\omega_{\mathrm{p}}t}-ae^{i\omega_{\mathrm{p}}t}\right),
\end{split}
\end{align}
where $a$ and $b$ are the bosonic operators for the pump and Stokes modes,
$R$ represents the Raman phonon mode. $g_{\mathrm{R}}$ is the nonlinear coupling
strength of Raman scattering, $\varepsilon_{\mathrm{p}}=\sqrt{\frac{P_{\mathrm{in}}}{\hbar\omega_{\mathrm{p}}}}$
is the driving strength. In the rotating frame of $\omega_{\mathrm{p}}a^{\dagger}a+\omega_{b}b^{\dagger}b+\left(\omega_{\mathrm{p}}-\omega_{b}\right)R^{\dagger}R$,
the cavity field in mode $a$ can be written as the sum of the mean
field $\alpha$ and the operator $a$, $a=\alpha+a$. By neglecting
the fluctuation in mode $a$, we have
\begin{gather}
\frac{d}{dt}\alpha=\left(-i\delta_{a}-\kappa_{a}/2\right)\alpha-ig_{\mathrm{R}}bR+\sqrt{\kappa_{\mathrm{e},a}}\varepsilon_{\mathrm{p}},\label{S2}\\
\frac{d}{dt}b=(-\kappa_{b}/2)b-ig_{\mathrm{R}}\alpha R^{\dagger}+\sqrt{\kappa_{b}}b_{\mathrm{in}},\label{S2}\\
\frac{d}{dt}R=\left(-i\delta_{\mathrm{R}}-\gamma_{\mathrm{R}}/2\right)R-ig_{\mathrm{R}}\alpha b^{\dagger}+\sqrt{\gamma_{\mathrm{R}}}R_\mathrm{in}\label{S2},
\end{gather}
where $b_{\mathrm{in}}$ is the input noise due to its coupling with the environment
modes and the same for the phonon mode $R$. Transform the differential equations
to the frequency domain by
\begin{gather}
o\left(\omega\right)=\int dto\left(t\right)e^{i\omega t},\\
o^{\dagger}\left(-\omega\right)=\int dto^{\dagger}\left(t\right)e^{i\omega t},
\end{gather}
we obtain
\begin{gather}
\left[i\omega-\kappa_{b}/2\right]b\left(\omega\right)-ig_{\mathrm{R}}\alpha R^{\dagger}\left(-\omega\right)+\sqrt{\kappa_{b}}b_{\mathrm{in}}\left(\omega\right)=0,
\end{gather}
\begin{flalign}
\begin{split}
\left[-i\left(\delta_{\mathrm{R}}-\omega\right)-\gamma_{\mathrm{R}}/2\right]R\left(\omega\right)-ig_{\mathrm{R}}\alpha b^{\dagger}\left(-\omega\right)\\+\sqrt{\gamma_{\mathrm{R}}}R_{\mathrm{in}}\left(\omega\right)=0,
\end{split}
\end{flalign}
where
\begin{eqnarray*}
\alpha\left(\omega\right)=\frac{ig_{\mathrm{R}}\int dte^{i\omega t}b\left(t\right)R\left(t\right)-\sqrt{\kappa_{\mathrm{e},a}}\varepsilon_{\mathrm{p}}}{-i\left(\delta_{a}-\omega\right)-\kappa_{a}/2}\delta\left(\omega\right)\\
=\frac{ig_{\mathrm{R}}b\star R\left(\omega\right)-\sqrt{\kappa_{\mathrm{e},a}}\varepsilon_{\mathrm{p}}}{-i\left(\delta_{a}-\omega\right)-\kappa_{a}/2}\delta\left(\omega\right)
\end{eqnarray*}
contains the backaction from modes $b$ and $R$. The solution of the
equations is 
\begin{gather}
b\left(\omega\right)=-\frac{\sqrt{\kappa_{b}}b_{\mathrm{in}}\left(\omega\right)+ig_{\mathrm{R}}\alpha\frac{\sqrt{\gamma_{\mathrm{R}}}R_{\mathrm{in}}^{\dagger}\left(-\omega\right)}{\alpha_{\mathrm{R}}^{-}}}{\alpha_{b}^{+}-\frac{g_{\mathrm{R}}^{2}|\alpha|^{2}}{\alpha_{\mathrm{R}}^{-}}},\\
R\left(\omega\right)=-\frac{\sqrt{\gamma_{\mathrm{R}}}R_{\mathrm{in}}\left(\omega\right)+ig_{\mathrm{R}}\alpha\frac{\sqrt{\kappa_{b}}b_{\mathrm{in},0}^{\dagger}\left(-\omega\right)}{\alpha_{b}^{-}}}{\alpha_{\mathrm{R}}^{+}-\frac{g_{\mathrm{R}}^{2}|\alpha|^{2}}{\alpha_{b}^{-}}},\label{eq:sol}
\end{gather}
where $\alpha_{b}^{+}=i\omega-\kappa_{b}/2=\alpha_{b}^{-}$, $\alpha_{\mathrm{R}}^{+}=-i\left(\delta_{\mathrm{R}}-\omega\right)-\gamma_{\mathrm{R}}/2$,
$\alpha_{\mathrm{R}}^{-}=i\left(\delta_{\mathrm{R}}+\omega\right)-\gamma_{\mathrm{R}}/2$. The
convolution term is 
\begin{eqnarray}
a\star b\left(\omega\right)\delta\left(\omega\right) =\int d\omega'a\left(\omega'\right)b\left(\omega-\omega'\right)\delta\left(\omega\right)\nonumber \\
=\int d\omega'a\left(\omega'\right)b\left(-\omega'\right).\label{eq:back}
\end{eqnarray}
The power spectrum of the intracavity field of mode $b$ is derived
as
\begin{gather}
S_{b}\left(\omega\right)=\langle b^{\dagger}\left(\omega\right)b\left(\omega\right)\rangle\nonumber=\\\nonumber\frac{\kappa_{b}\langle b_{\mathrm{in}}^{\dagger}\left(\omega\right)b_{\mathrm{in}}\left(\omega\right)\rangle}{|\alpha_{b}^{+}-\frac{g_{\mathrm{R}}^{2}|\alpha|^{2}}{\alpha_{\mathrm{R}}^{-}}|^{2}}+\frac{\frac{g_{\mathrm{R}}^{2}|\alpha|^{2}}{|\alpha_{\mathrm{R}}^{-}|^{2}}\kappa_{b}\langle R_{\mathrm{in}}\left(-\omega\right)R_{\mathrm{in}}^{\dagger}\left(-\omega\right)\rangle}{|\alpha_{b}^{+}-\frac{g_{\mathrm{R}}^{2}|\alpha|^{2}}{\alpha_{\mathrm{R}}^{-}}|^{2}}\\=\nonumber\\
\frac{\kappa_{b}|\alpha_{\mathrm{R}}^{-}|^{2}\langle b_{\mathrm{in}}^{\dagger}\left(\omega\right)b_{\mathrm{in}}\left(\omega\right)\rangle}{|\alpha_{b}^{+}\alpha_{\mathrm{R}}^{-}-g_{\mathrm{R}}^{2}|\alpha|^{2}|^{2}}+\frac{g_{\mathrm{R}}^{2}|\alpha|^{2}\gamma_{\mathrm{R}}\langle R_{\mathrm{in}}\left(-\omega\right)R_{\mathrm{in}}^{\dagger}\left(-\omega\right)\rangle}{|\alpha_{b}^{+}\alpha_{\mathrm{R}}^{-}-g_{\mathrm{R}}^{2}|\alpha|^{2}|^{2}}\nonumber\\
=\frac{1}{2\pi}\frac{\gamma_{\mathrm{R}}g_{\mathrm{R}}^{2}|\alpha|^{2}\left(n_{th}+1\right)}{|\alpha_{b}^{+}\alpha_{\mathrm{R}}^{-}-g_{\mathrm{R}}^{2}|\alpha|^{2}|^{2}}
\end{gather}
where $n_{th}$ is the mean phonon number in the thermal reservoir
of the phonon mode. The Raman threshold appears at the frequency where
$S_{a}\left(\omega\right)$ is the largest. $\alpha_{b}^{+}\alpha_{\mathrm{R}}^{-}-g_{\mathrm{R}}^{2}|\alpha|^{2}\rightarrow0$
gives
\begin{eqnarray}
-\omega\left(\delta_{\mathrm{R}}+\omega\right)+\kappa_{b}\gamma_{\mathrm{R}}/4 & = & g_{\mathrm{R}}^{2}|\alpha|^{2},\\
\omega\gamma_{\mathrm{R}}+\left(\delta_{\mathrm{R}}+\omega\right)\kappa_{b} & =0 & ,
\end{eqnarray}
Therefore, the lasing frequency relative to the resonant frequency of the optical mode $b$ is 
\begin{eqnarray*}
\omega_{s} & = & -\frac{\delta_{\mathrm{R}}\kappa_{b}}{\gamma_{\mathrm{R}}+\kappa_{b}},
\end{eqnarray*}
and the lasing threshold is
\begin{eqnarray}
|\alpha_{th}|^{2} & = & \frac{\kappa_{b}\gamma_{\mathrm{R}}}{4g_{\mathrm{R}}^{2}}\left[1+\frac{(2\delta_{\mathrm{R}})^{2}}{\left(\gamma_{\mathrm{R}}+\kappa_{b}\right)^{2}}\right].\label{eq:th1}
\end{eqnarray}
For $\kappa_{\mathrm{R}}\gg\kappa_{b}$, the threshold reduces to
\begin{eqnarray*}
|\alpha_{th}|^{2} & = & \frac{\kappa_{b}\gamma_{\mathrm{R}}}{4g_{\mathrm{R}}^{2}}\left[1+\frac{(2\delta_{\mathrm{R}})^{2}}{\gamma_{\mathrm{R}}^{2}}\right].
\end{eqnarray*}
which corresponds Eq. (1) in the main text, where the symbol $\kappa_{b}$ is replaced by $\kappa_{\mathrm{R}}$ in the main text with the subscript $\mathrm{R}$ referring to the Stokes mode. Moreover, the pump field energy will be clamped at $\hbar\omega_\mathrm{p}\gamma_\mathrm{R}\kappa_{b}[1+(\frac{2\delta_\mathrm{R}}{\gamma_{R}})^2]/{4g^2_\mathrm{R}}$ once the onset of $1^{\mathrm{st}}$ Raman lasing is reached.
\\
\\
\\
\section{Experimental assessment of of $g_{\mathrm{R}}$} 

\begin{figure}
\begin{centering}
\includegraphics[width=1\columnwidth]{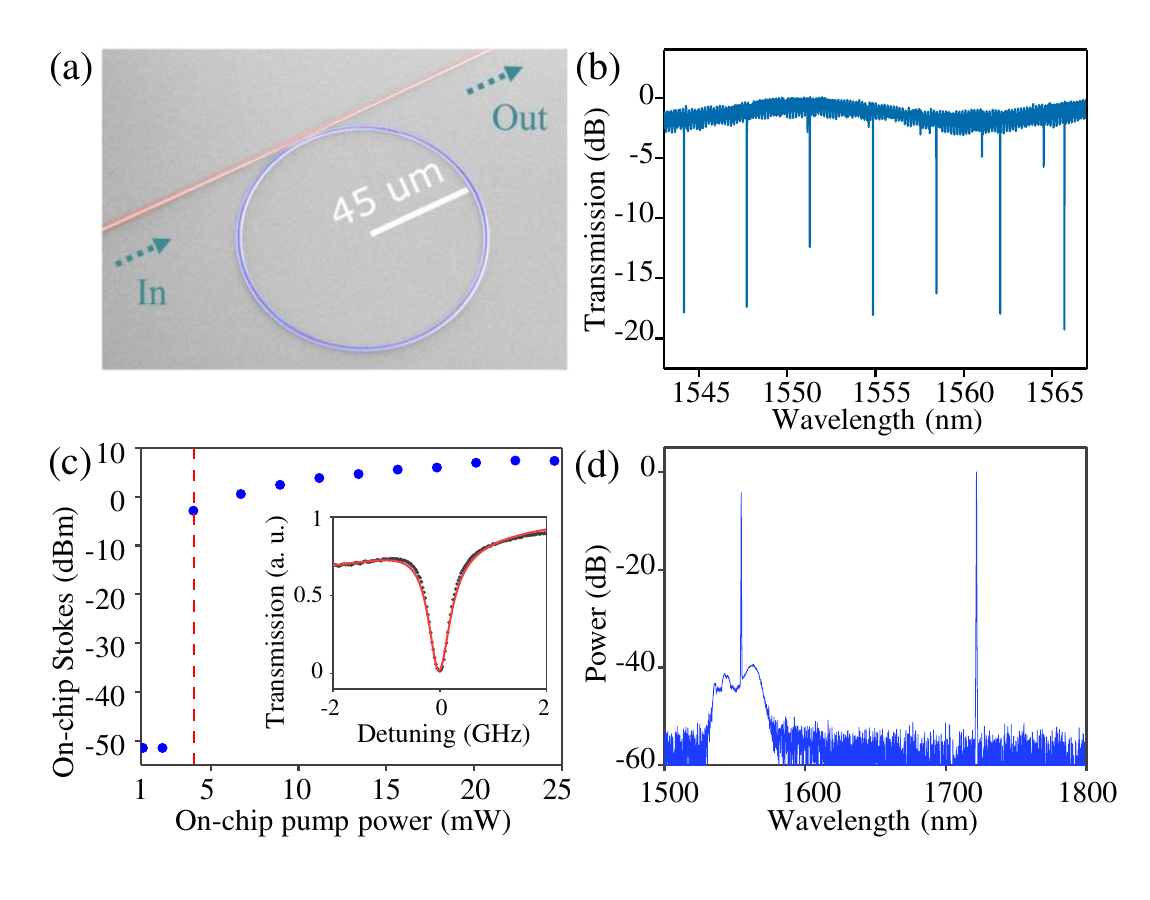}
\par\end{centering}
\caption{\label{fig:S3} (a) False-color scanning electron micrograph (SEM) of a single waveguide couled LN microring. (b) TE transmission of the microring. (c) On-chip Stokes light power versus on-chip pump power. Inset: zoom-in view of the TE00 pump resonance. (d) An optical spectrum of the Raman lasing under an on-chip input power of $\sim20$\,mW.}
\end{figure}

In the experiment, we assess the $1^{\mathrm{st}}$ Raman lasing threshold and $g_R$ in a conventional single waveguide coupled LN microring whose dimension is identical to the U-ring in Fig. 3 of the main text. The SEM of the microring is shown in Fig. \ref{fig:S3}(a), and TE-transmission of the micoring is partly presented in Fig. \ref{fig:S3}(b). 
Fig. \ref{fig:S3}(c) plots the maximum $1^{\mathrm{st}}$ Stokes power under gradually increased pump power swept across the TE00 mode at 1554.4,nm. The threshold of the $1^{\mathrm{st}}$ Raman is found to be $\sim4$\,mW. As an example, a normalized Raman lasing spectrum is shown in Fig. \ref{fig:S3}(d) under $\sim20$\,mW on-chip input power which exhibits a Raman shift of 625\,cm$^{-1}$, indicating the Stokes field almost overlaps with the E(LO$_8$) Raman gain center ($\delta_\mathrm{R}\approx0$) \cite{Ridah_1997_2}. A zoom-in view of the pump resonance is presented in the inset of Fig. \ref{fig:S3}(c) with the external and intrinsic coupling rates extracted to be $\kappa_{\mathrm{e}}/(2\pi)=280$\,MHz and $\kappa_{\mathrm{i}}/(2\pi)=220$\,MHz respectively. $g_{R}$ then can be calculated from the measured Raman lasing threshold based on Eq. 1 in the main text, which is estimated to be 1.1$\times10^{7}$\,rad assuming the Stokes mode have similar $\kappa_{\mathrm{e}}$ and $\kappa_{\mathrm{i}}$ as the pump mode under this single-straight-waveguide coupling configuration.

\begin{figure}
\begin{centering}
\includegraphics[width=1\columnwidth]{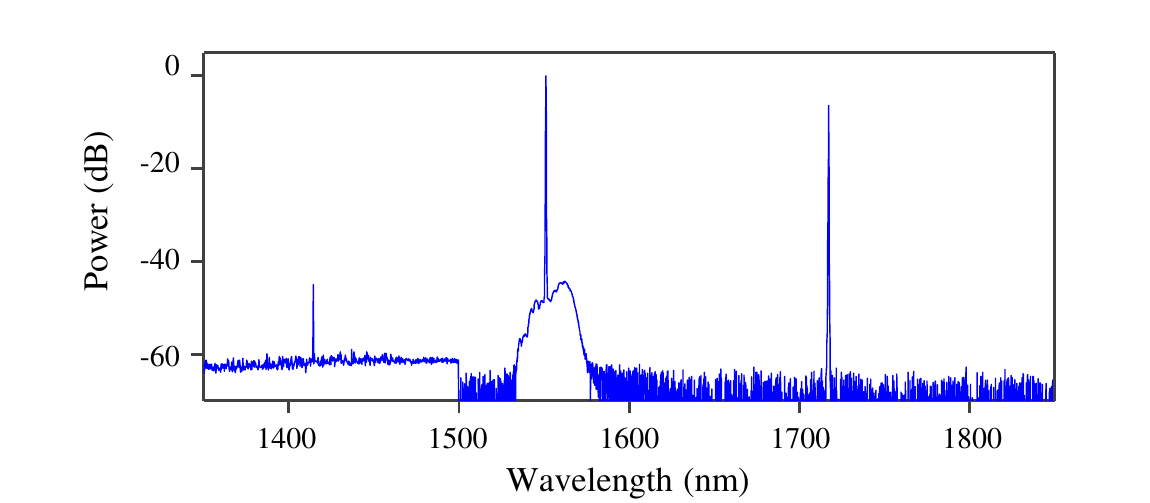}
\par\end{centering}
\caption{\label{fig:S4} An optical spectrum of Raman lasing from the LN microring shown in Fig. \ref{fig:S3}(a), which dominants the cavity dynamics.}
\end{figure}

For this device (Fig. \ref{fig:S3}(a)), without reservoir engineering implemented, the Raman lasing threshold $\varepsilon_\mathrm{R,th}$ is 2.2 times lower than that of reservoir engineered device shown in the main text. Our analytical calculation shows $(\varepsilon_\mathrm{R,th}-\varepsilon_\mathrm{S,th})/\varepsilon_\mathrm{S,th}=-0.1$ assuming $\kappa_{\mathrm{e}}/(2\pi)=280$\,MHz and $\kappa_{\mathrm{i}}/(2\pi)=220$\,MHz for all modes of interest. And indeed, Raman lasing (Fig. \ref{fig:S4}) is found to be dominant and no soliton combs are observed in the experiment, consistent with our analysis.
\\
\section{Net external coupling rate of the U-ring}
\begin{figure}[h]
\begin{centering}
\includegraphics[width=0.5\columnwidth]{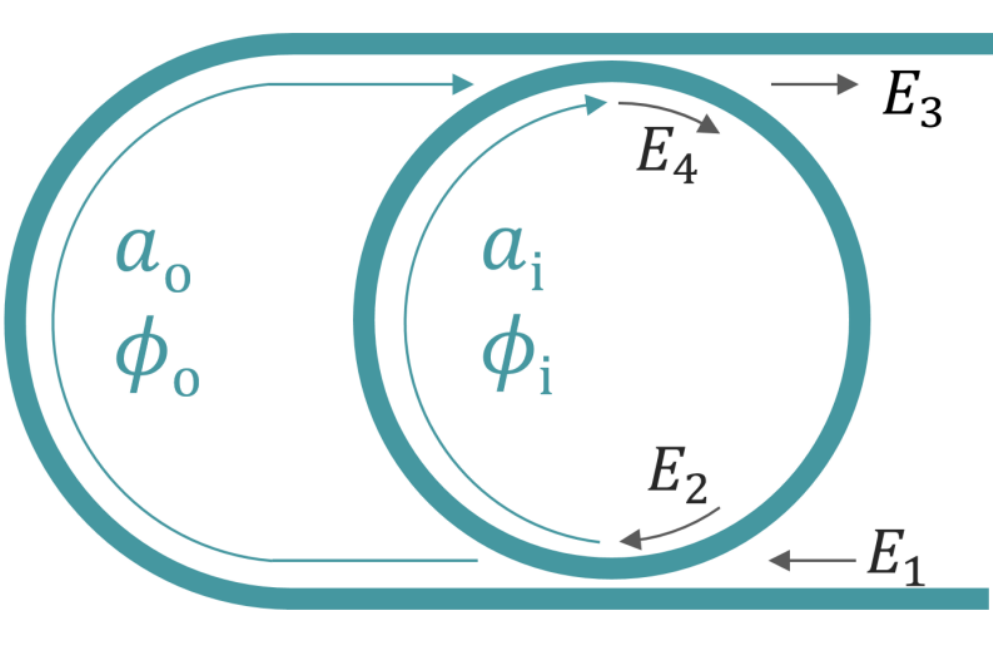}
\par\end{centering}
\caption{\label{fig:S5} Schematics of the U-ring. $E_{1}$ ($E_{3}$) refers to the input (output) optical field, and $E_{2}$ ($E_{4}$) is the intracavity field immediately before (after) the lower (upper) microring-waveguide coupler. $a_{\mathrm{i}(\mathrm{o})}$ and $\phi_{\mathrm{o}(\mathrm{o})}$ are the optical field amplitude transmittance and linear phase accumulation, respectively, along the inner-arm/half-ring (outer arm/U-shaped-arm) between the two microring-waveguide coupling points.}
\end{figure}

The schematics of the U-ring is depicted in Fig. \ref{fig:S5}. And the optical fields within different sections of the waveguides are related to each other by 
\begin{gather}
\begin{bmatrix}
    E_{3}      \\
    E_{4}      
\end{bmatrix}
= 
\begin{bmatrix}
    r  &  it      \\
    it  &  r      
\end{bmatrix} 
\begin{bmatrix}
    a_\mathrm{o}e^{i\phi_{\mathrm{o}}}  &  0      \\
    0  &  a_\mathrm{i}e^{i\phi_{\mathrm{i}}}      
\end{bmatrix}
\begin{bmatrix}
    r  &  it      \\
    it  &  r      
\end{bmatrix} 
\begin{bmatrix}
    E_{1}      \\
    E_{2}      
\end{bmatrix}
\label{S8}\\
    E_{2}=a_\mathrm{i}e^{i\phi_{\mathrm{i}}}E_{4}
\label{S9}    
\end{gather}
\noindent
where $r$ and $t$ represent the optical field self- and cross- coupling coefficients, respectively, of the two identical microring-waveguide couplers and satisfy $r^2+t^2=1$ \cite{heebner2008optical} neglecting parametric loss. The value of $r$ relates to the  single waveguide external coupling rate $\kappa_{0}$ as $r^{2}=e^{-\kappa_{0}t_\mathrm{R}}$ with $t_\mathrm{R}$ denoting the microring roundtrip time, and similarly the transmittance in the inner reference arm ($a_\mathrm{i}$) follows $a^{2}_{\mathrm{i}}=e^{-\kappa_\mathrm{i}t_\mathrm{R}/2}$. Assuming $a_{\mathrm{o}}\approx{a_{\mathrm{i}}\approx1}$ and $t^2\ll{r^2}\approx{1-\kappa_{0}t_\mathrm{R}}$, then the overall transmission in a resonance can be derived as
\begin{gather}
\left|\frac{E_{3}}{E_{1}}\right|^2=\left|\frac{2\kappa_{\mathrm{0}}(1+cos\Delta{\phi_{\mu}})-\kappa_{\mathrm{i}}}{2\kappa_{\mathrm{0}}(1+cos\Delta{\phi_{\mu}})+\kappa_{\mathrm{i}}}\right|^2
\label{S10}
\end{gather}
\noindent
with the $\Delta\phi_\mu=\phi_{\mathrm{i}}-\phi_{\mathrm{o}}$ representing the phase difference between the inner arm (half-ring) and the outer arm (U-shaped arm) for the cavity mode $\mu$. Thus, the net external coupling rate follows 
\begin{gather}
\kappa_{\mathrm{e},\mu}=2\kappa_{\mathrm{0}}(1+cos\Delta{\phi_{\mu}})
\label{S11}
\end{gather}
\noindent
which corresponds to Eq. 3 in the main text. 
\\
\\
\\
\emph{Experiment setup for accessing soliton state} 
\\
\par
\begin{figure}
\begin{centering}
\includegraphics[width=1\columnwidth]{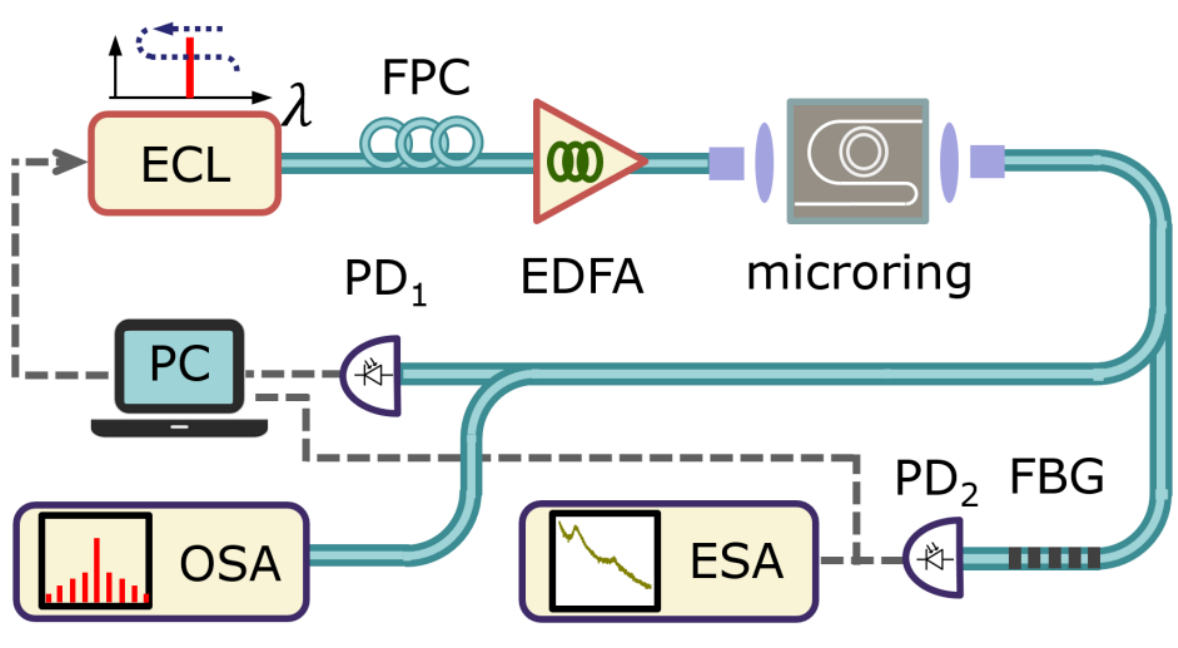}
\par\end{centering}
\caption{\label{fig:S6} Schematics of the experiment setup. ECL, external cavity laser; FPC, fiber polarization controller; EDFA, erbium-doped fiber amplifier; FBG, fiber Brag gating; PD, photodetector; ESA, electrical spectrum analyzer; OSA, optical spectrum analyzer; PC, personal computer, which is used to control the laser scanning as well as to record the transmission and comb power. Optical fibers and electrical cables are presented by solid cyan lines and dashed gray lines respectively.}
\end{figure}

The experiment setup is schematically shown in Fig. \ref{fig:S6}. The output wavelength $\lambda_{\mathrm{c}}$ of the external cavity laser (ECL) can be programmatically scanned across microring resonances. An erbium-doped fiber amplifier (EDFA) is used to amplify the ECL's output to pump the U-ring, and a fiber polarization controller (FPC) is employed to adjust the polarization state of the pump field before coupled into the chip. Subsequently, the output of the chip is recorded and analyzed with direct photodetection (PD1), an optical spectrum analyzer (OSA) and a photodetector (PD2) followed by an electrical spectrum analyzer (ESA), respectively. To monitor comb power, a fiber Bragg grating (FBG) is used to suppress the pump component of the comb before the photodetector PD2.

To characterize the dispersion of the U-ring resonator used for the experiment (Fig. 3 of the main text), we measure the integrated dispersion \cite{PhysRevLett.113.123901} around the pump mode at 1554.4\,nm. By fitting the experiment data, the second order dispersion is extracted to be $D_{2}/(2\pi)\approx{20}$\,MHz. 
\begin{figure}[h]
\begin{centering}
\includegraphics[width=1\columnwidth]{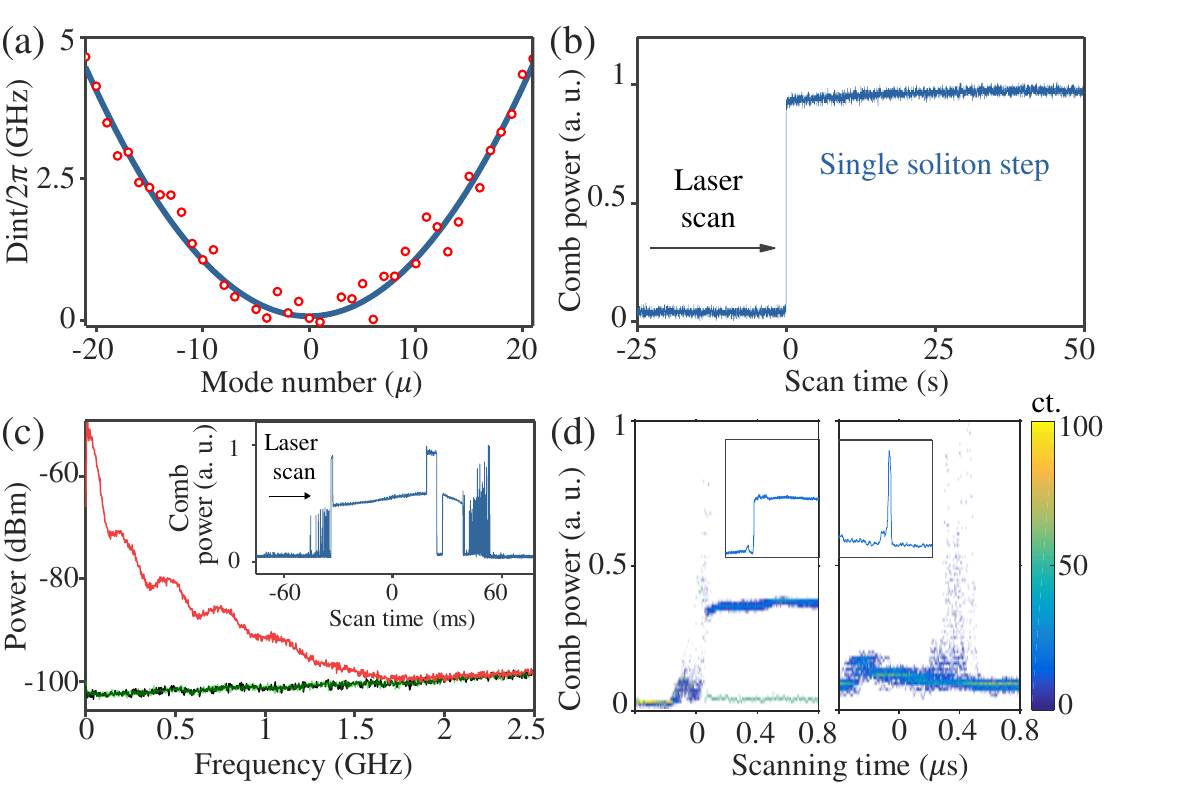}
\par\end{centering}
\caption{\label{fig:S7} (a) The measured (red dots) integrated dispersion of the device with parabola-fitting curve (blue line). (b) Comb power trace under relatively slow red-to-blue laser scan (83\,MHz/s). (c) Relative intensity noise spectra of MI comb state (red), soliton comb (green) and the detector background (black). Inset: one example of comb power trace under red-to-blue laser scan (62.5\,GHz/s). (d) Overlaid comb power traces for $\varepsilon_\mathrm{S,th}<\varepsilon_\mathrm{R,th}$ (left) and $\varepsilon_\mathrm{S,th}>\varepsilon_\mathrm{R,th}$ (right) under pump scans at 1\,GHz/$\mu$s. Color bar scales with the pixel counts. Insets: single shots of soliton step generation and Raman lasing under laser scan.}
\end{figure}
To access the soliton state in the case where the pump mode is selected such that $\varepsilon_\mathrm{S,th}<\varepsilon_\mathrm{R,th}$, we scan the pump into the resonance from the red-detuned regime until spontaneous soliton mode-locking occurs under the LN's photorefractive effect \cite{He:19,Gong:19}. For example, when we scan the pump toward the resonance from its red-side at a speed of 83\,MHz/s, the cavity dynamics spontaneously evolves into a single soliton state through the self-start of mode-locking process induced by photorefractive effect \cite{He:19}, as manifested in the measured comb power trace shown in Fig. \ref{fig:S7}(b). Then, by stopping the laser scan, soliton comb can be steadily obtained. When the pump is scanned at faster speed (e. g. 62.5\,GHz/s) across the resonance from red to blue, the comb power trace exhibits more chaotic dynamics (Fig. \ref{fig:S7}(c) inset) under the photorefractive effect \cite{He:19} with soliton steps revealed amid other noisy comb states. The relative intensity noise spectra of a MI comb, a soliton comb and detector background are presented in Fig. \ref{fig:S7}(c) for comparison, indicating the low noise feature of the soliton state.  

Lastly we experimentally investigate the soliton generation statistics in the U-ring when pumped at different resonances, by counting the occurrence of soliton steps over multiple laser scans. To facilitate the process, we use an external single-side-band modulator \cite{PhysRevApplied.9.024030,Gong:18} to sweep the pump wavelength across the resonances periodically and record the comb power traces simultaneously. When the pump is scanned across the mode at 1554.4\,nm for which the Raman threshold is lifted ( $\varepsilon_\mathrm{S,th}<\varepsilon_\mathrm{R,th}$), we obtain a $\sim$\,50$\%$ success rate of launching a single soliton state as indicated from the overlaid comb power traces shown in the left plot of Fig. \ref{fig:S7}(d). However, when we scan across the resonance at 1558\,nm where $\varepsilon_\mathrm{S,th}>\varepsilon_\mathrm{R,th}$, no soliton steps are observed other than the Raman lasing signals. The statistics reveals a from-0$\%$-to-50$\%$ change in the soliton generation success rate as $\varepsilon_\mathrm{S,th}$ is made smaller than $\varepsilon_\mathrm{R,th}$, confirming the effectiveness of dissipation control in suppressing Raman effects for soliton generation. We attribute the non-unity success rate, in part, to the fact that there may be finite \enquote{no-soliton-steps} probability even in the Kerr effect dominated scenario \cite{universaldynamics,Gong:18}.

\section{Cascaded interference couplers} 

\begin{figure}[h]
\begin{centering}
\includegraphics[width=1\columnwidth]{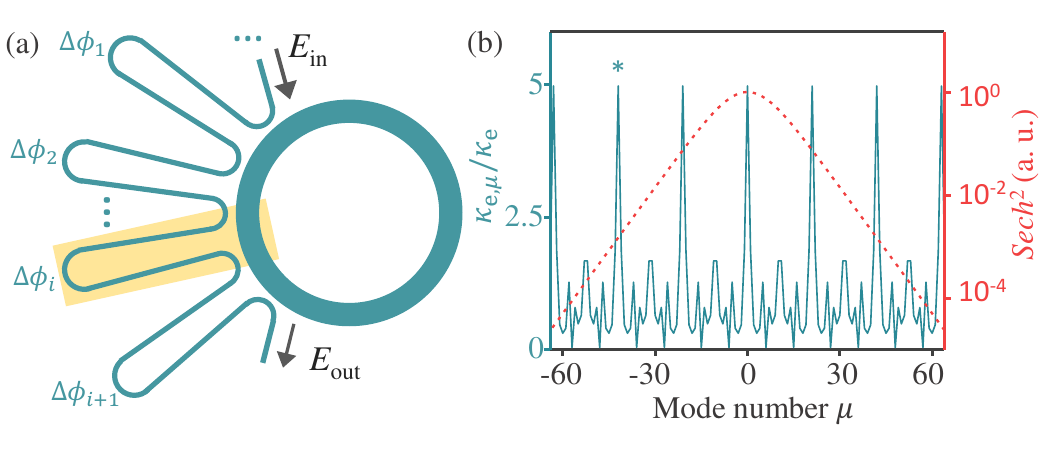}
\par\end{centering}
\caption{\label{fig:S8} (a) The scheme of cascaded self-interferencing on the same LN microring as that in Fig. 3 of the main text. $\Delta{\phi_i}$ refers to the phase-delay difference of the $i^{\mathrm{th}}$ interferometer. Yellow shaded area outlines one of the cascaded interferometers. (b) The calculated $\kappa_{\mathrm{e},\mu}/\kappa_{\mathrm{e}}$ (cyan) of the 4-arm cascaded configureation, with a normalized $sech^{2}-$shaped intracavity soliton energy envelop with a FWHM of 6.5\,THz as an example (red, in log-scale). Here, the external coupling rate at each microring-waveguide coupling point is set as $\kappa_{0}/(2\pi)=100$\,MHz such that the $sech^{2}-$weighted $\kappa_{\mathrm{e}}/(2\pi)$ remains the same (480\,MHz) to the one in the experiment. (b) The asterisk marks the Stokes mode ($\mu=-$42) that overlaps with the Raman center and also sees the peak value of $\kappa_{\mathrm{e},\mu}$. $\mu=0$ is the pump mode.}
\end{figure}

It is possible to further enhance our system's ability to suppress stronger Raman effect if desired. For example, a series of U-arms can be cascaded along the microring, as schematically depicted in Fig. \ref{fig:S8}(a), in an effort to enlarge the ratio of $(\varepsilon_\mathrm{R,th}-\varepsilon_\mathrm{S,th})/\varepsilon_\mathrm{S,th}$.
When there are 4 U-arms cascaded, the system's net coupling rate becomes (assuming $\kappa_{\mathrm{0}}t_{R}\ll1$)
\begin{flalign}
\begin{split}
\kappa_{\mathrm{e},\mu}=\kappa_{\mathrm{0}}(5+2\sum_{i=1}^{4}cos\Delta{\phi_{i,\mu}}+2\sum_{i=1}^{3}cos(\Delta{\phi_{i,\mu}}+\Delta{\phi_{i+1,\mu}})\\+2\sum_{i=1}^{2}cos(\Delta{\phi_{i,\mu}}+\Delta{\phi_{i+1,\mu}}+\Delta{\phi_{i+2,\mu}})\\+2\sum_{i=1}^{1}cos(\Delta{\phi_{i,\mu}}+\Delta{\phi_{i+1,\mu}}+\Delta{\phi_{i+2,\mu}}+\Delta{\phi_{i+3,\mu}}))
\label{S12}
\end{split}
\end{flalign}
\noindent
and is plotted in Fig. \ref{fig:S8}(b) at each soliton-forming mode, where $\Delta{\phi}_{i,\mu}=\omega_{\mu}\Delta{L}_{i}/c$, with $\omega_{\mu}$, $\Delta{L}_{i}=L^{i}_\mathrm{o}n^{i}_\mathrm{o}-L^{i}_\mathrm{i}n^{i}_\mathrm{i}$ and $c$ denoting the ${\mu}^{\mathrm{th}}$ microring resonance angular frequency, optical path length difference of the $i^{\mathrm{th}}$ interferometer and speed of light in vacuum respectively. Here, the arm length $\Delta{L}_{i}$ is set as $\Delta{L}_{1}=\Delta{L}_{2}/2=\Delta{L}_{3}/2=\Delta{L}_{4}/4=c/(21\mathrm{FSR})$. For simplicity, $\Delta{L}_{i}$ is assumed to be invariant over the frequency range of interest. As can been seen from Fig. \ref{fig:S8}(b), this cascaded modulation of $\kappa_{\mathrm{e,\mu}}$  features reduced duty-cycle, imposes stronger $\kappa_{\mathrm{e,R}}$ at the Stokes mode while keeping much lower $\kappa_{\mathrm{e,\mu}}$ for the rest of soliton-forming modes and gives rise to an enhanced dynamic range in controlling external coupling rates i.e. $\kappa_{\mathrm{e,R}}/\kappa_{\mathrm{e}}$ ($\kappa_{\mathrm{e}}$ is the $sech^{2}-$weighted external coupling rate). Applying this modulation scheme to the same microring in Fig. 3 of the main text, greater value of $(\varepsilon_\mathrm{R,th}-\varepsilon_\mathrm{S,th})/\varepsilon_\mathrm{S,th}=1.6$ can be achieved, which is a 16-time improvement, based on the estimation using weighted $\kappa_{\mathrm{e}}$ to generate a soliton with the same target FWHM. In other words, this 4-cascade modulation scheme, if realized, can compensate stronger Raman coupling rate up to ${1.6}g_{\mathrm{R}}$ for such soliton generation. 

\begin{figure}[h]
\begin{centering}
\includegraphics[width=0.8\columnwidth]{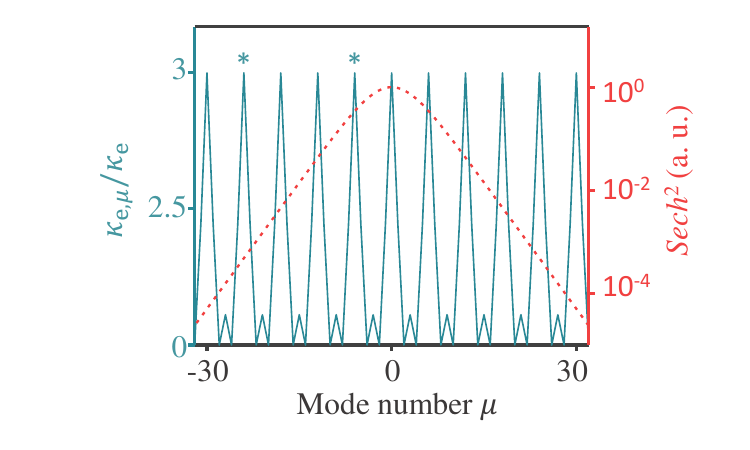}
\par\end{centering}
\caption{\label{fig:S9}Calculated $\kappa_{\mathrm{e},\mu}/\kappa_{\mathrm{e}}$ (cyan curve) of a 2 U-arm-cascaded LN microring along with a normalized $sech^{2}-$shaped intracavity soliton energy envelop with a FWHM of 6.5\,THz as an example (red, in log-scale). Here, the external coupling rate at each microring-waveguide coupling point is set as $\kappa_{0}/(2\pi)=160$\,MHz such that the $sech^{2}-$weighted $\kappa_{\mathrm{e}}/(2\pi)$ remains the same (480\,MHz) to the one in the experiment. Asterisks mark the Stokes modes at $\mu\approx$\,-24 and -6 that overlap with the Raman gain centers of $\mathrm{E}(\mathrm{LO}_{8})$, $\mathrm{E}(\mathrm{TO}_{1})$ Raman-active phonons in LN \cite{Ridah_1997_2}. $\mu=0$ is the pump mode.}
\end{figure}

Additionally, in case of multiple dominant Raman gain centers overlapping with the soliton forming modes, we may also be able to take advantage of the periodic modulation of $\kappa_{\mathrm{e},\mu}$ based on the self-interference structures to suppress them all. For example, we consider the case of a LN microring with a FSR\,$=$\,775\,GHz whose soliton forming modes overlap with two Raman gain centers at $\mu\approx$\,-24 and -6, as shown in Fig. \ref{fig:S9}, where one Raman gain center is much closer to the pump mode than the other and both of them are assumed to be the dominant in this microring. To impose higher $\kappa_{\mathrm{e},\mu}$ at the affected modes, we can cascade 2-interferometer along this microring and set the modulation period close to a common divisor of the target two Raman phonon frequencies, which translates to the lengths of the two arms to be designed as $\Delta{L}_{1}=\Delta{L}_{2}=c/(6\mathrm{FSR})$ as an example for this case. By doing so, we can align both the affected soliton forming modes to the peak of $\kappa_{\mathrm{e},\mu}$ curve to suppress Raman lasings.

\begin{figure}[h]
\begin{centering}
\includegraphics[width=1\columnwidth]{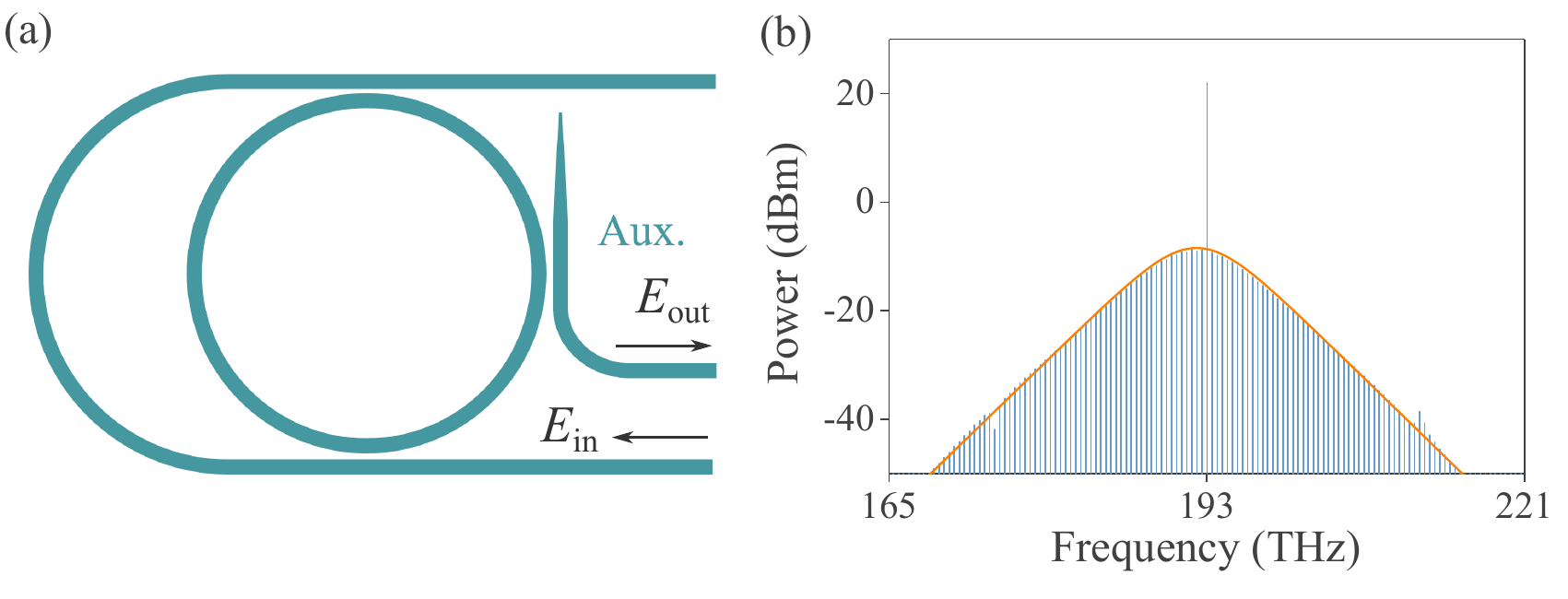}
\par\end{centering}
\caption{\label{fig:S10}\textcolor{black}{(a) Scheme for drop-port extraction of the U-ring's output via an auxiliary point-coupled waveguide. (b) The simulated output spectrum. The spectrum exhibits the predicted $sech^{2}$-shaped profile from a soliton. The device parameters are the same as those used in Fig. \ref{fig:S1}, and the auxiliary waveguide has a coupling rate of $\kappa_{\mathrm{aux}}$\,=\,50\,MHz as an example.}}
\end{figure}

\textcolor{black}{In certain applications, a smooth soliton power spectrum is desired. This can be achieved by introducing a drop-port to the U-ring to sample a smoother soliton spectrum. The schematics and simulated output spectrum are presented in Fig. \ref{fig:S10}. Note the external coupling rates become $\kappa_{\mathrm{e},\mu}$\,=$2\kappa_{\mathrm{0}}(1+cos\Delta{\phi_{\mu}})+\kappa_{\mathrm{aux}}$, where is $\kappa_{\mathrm{aux}}$ the coupling rate of the added waveguide.}

Based on above discussion, we believe our dissipation control capacity can be further extended with more complex interferometers and design of arm lengths to compensate for multiple strong Raman gains. And our work will also inspire and provide guidance for other forms of dissipation control, not only confined with the self-interference scheme in our case.

\bibliographystyle{apsrev4-1}

\end{document}